\documentclass[a4paper,11pt]{article}
\pdfoutput=1 

\usepackage{jheppub} 

\usepackage[T1]{fontenc} 

\usepackage{xcolor,comment}

\usepackage[textsize=tiny]{todonotes}

\usepackage{braket, bm}

\title{\boldmath Edge modes and dressing fields for the Newton-Cartan quantum Hall effect}

\author[a,1]{William J. Wolf,\note{Corresponding author.}}
\author[a]{James Read,}
\author[b]{Nicholas J. Teh}

\affiliation[a]{Faculty of Philosophy, University of Oxford,  \\Radcliffe Humanities, Woodstock Road, OX2 6GG, United Kingdom}
\affiliation[b]{
Department of Philosophy, University of Notre Dame, \\ Notre Dame, IN, 46556,
United States}

\emailAdd{william.wolf@philosophy.ox.ac.uk}
\emailAdd{james.read@philosophy.ox.ac.uk}
\emailAdd{nteh@nd.edu}

\abstract{It is now well-known that Newton-Cartan theory is the correct geometrical setting for modelling the quantum Hall effect. In addition, in recent years edge modes for the Newton-Cartan quantum Hall effect have been derived. However, the existence of these edge modes has, as of yet, been derived using only orthodox methodologies involving the breaking of gauge-invariance; it would be preferable to derive the existence of such edge modes in a gauge-invariant manner. In this article, we employ recent work by Donnelly and Freidel in order to accomplish exactly this task. Our results agree with known physics, but afford greater conceptual insight into the existence of these edge modes: in particular, they connect them to subtle aspects of Newton-Cartan geometry and pave the way for further applications of Newton-Cartan theory in condensed matter physics.}

\begin{document} 
\maketitle
\flushbottom

\section{Introduction}

Understanding the empirical content of gauge symmetries has long constituted a major topic in the foundations of physics. Physical systems with boundaries are of particular interest when probing this question as the presence of boundaries can break gauge invariance; the ways in which this is accounted for can lead to the propagation of physical edge modes on the boundary. Recent work by Donnelly and Freidel has shown that we can account for these degrees of freedom on the boundary by enhancing the original symmetries of the system \cite{DL}. In particular, Donnelly and Freidel's insight was that edge modes can be identified through using the covariant Hamiltonian formalism and demanding that the theory exhibit gauge invariance at the boundary through introducing new boundary fields. This effectively allows one to systematically identify and investigate edge modes and sheds light on the empirical significance of the system's symmetries. 

The quantum Hall effect (QHE) is one such system that exhibits this kind of bulk-boundary correspondence of interest. The QHE refers to an effect that systems of electrons exhibit when confined to two dimensions at low temperatures and subjected to an external magnetic field (see \cite{Tong} for a review). Among the surprises is that the electrical conductivity of such a system takes quantised values that plateau in proportion to $e^2/h$ and are functions of the magnetic field strength. These values can either be integers (in the case of the integer quantum Hall effect) or fractions (in the case of the fractional quantum Hall effect). Furthermore, these systems exhibit striking behavior on their boundaries in the form of edge modes. These edge modes encode many of the system's relevant physical properties and also feature prominently in the construction of conformal field theories using the Wess-Zumino-Witten (WZW) model, where these conformal field theories can then be used to further describe properties of these systems, such as their transitions between different quantised states \cite{Zirn}.

While much of this behavior is well-established, recent work has demonstrated that there are additional features of these quantum Hall systems which go beyond these quantised conductivity values: these features include Hall viscosity and corrections to the conductivity from electromagnetic perturbations \cite{Wen, Avron, ReadRezayi, DamSon1}. Son has pointed out that the standard Chern-Simons theory used to model quantum Hall systems does not encode these additional features and, furthermore, does not respect the appropriate symmetries of the microscopic theory. This necessitates finding an improved description. In particular, he has shown that such a description can be developed using Newton-Cartan theory, which has been studied previously as a geometric reformulation of Newtonian gravity (see e.g. \cite{TehRead, TehNewton, GeraciePrabhu, Malament} and references therein), because Newton-Cartan theory has the same non-relativistic symmetries as those possessed by systems of non-relativistic electrons \cite{DamSon1}. This discovery has shown that this Newton-Cartan formalism is the correct way to model the QHE, in turn leading to a renewed interest in the subject, and opening the door to further applications in the study of non-relativistic holography and coupling matter to non-relativistic backgrounds (see e.g.~\cite{GeraciePrabhu, Jensen1}).


In the present article, we combine the insights of Donnelly and Freidel with those of Son in order to construct edge modes for this Newton-Cartan model of the QHE using the machinery developed by Donnelly and Freidel \cite{DL}. This will reveal that, rather than edge modes arising from constraints on gauge transformations at the boundary (this being the canonical explanation for their existence: see \cite{Tong}), edge modes in this construction are the result of a new boundary field of the Newton-Cartan effective field theory for QHE. Additionally, this will shed light on the empirical significance of the familiar $U(1)$ gauge symmetry of QHE models as well as that of Milne symmetry, which is a new symmetry of the problem arising from the Newton-Cartan background. Having accomplished this work, we will draw on recent literature surrounding the direct empirical significance of symmetries (as discussed in \cite{GW, Teh}) in addition to recent literature concerning the so-called dressing field method and its implications for the significance of gauge symmetries (see e.g.~\cite{Francois}), in order to make clear the conceptual upshots and insights afforded by this approach to the QHE.


The structure of this article is as follows. In \S\ref{s2}, we review Son's application of the Newton-Cartan geometry to the QHE and the typical way in which one constructs edge modes for the theory. In \S\ref{s3}, we review the role of Milne invariance in Newton-Cartan geometry and the subsequent reformulation of Son's bulk model. In \S\ref{s4}, we examine the Donnelly-Freidel programme for identifying boundary observables and apply this method to the Milne invariant model of the QHE. In \S\ref{s5}, we discuss how this model of the Newton-Cartan QHE and the subsequent application of the Donnelly-Freidel programme informs the empirical content of the symmetries relevant in the construction.

\section{Newton-Cartan geometry and the quantum Hall effect}\label{s2}

In this section, we review Son's original application of Newton-Cartan geometry to the quantum Hall effect (\S\ref{s2.1}), as well as the subsequent construction of edge modes based on Son's model (\S\ref{s2.2}).

\subsection{Son's bulk model}\label{s2.1}

Quantum Hall systems are many-electron systems confined to two dimensions at low temperatures and subject to external electromagnetic fields. They exhibit two surprising and important topological properties: (i) the Hall conductivity and (ii) the Hall shift \cite{GeraciePrabhu}, which are, respectively, 
\begin{equation} \label{NCRelations}
\begin{split}
\sigma_{ij} = \frac{\nu}{2 \pi} \,,
\qquad
Q = \nu (N_{\phi}+\mathcal{S}) \,.
\end{split}
\end{equation}
Here $\sigma_{ij}$ is the Hall conductivity, $\nu$ is the filling factor (which is either an integer or fraction and is related to Landau levels occupied by the electrons in the system),
$Q$ is the charge of the system, $N_{\phi}$ is the magnetic flux $B/2\pi$, and $\mathcal{S}$ is the shift. (We use units setting $e^2/\hbar = 1$.) The Hall conductivity is the result that the electrical conductivity of these many-electron systems plateaus at quantised values that can be either integers or fractions, depending on the magnetic field strength. The Hall shift is a more recent discovery, and is related to the Hall viscosity, which is a non-dissipative transport coefficient. This effect means that the effective charge of the quantum Hall system is offset by the shift $\mathcal{S}$. 

We can describe these topological properties using topological quantum field theories. The standard way of exploring quantum Hall physics is via Chern-Simons theory, and is given by an effective action with the form
\begin{equation} \label{CS action}
   S_{CS}[A] = \frac{\nu}{4 \pi} \int d^3 x \epsilon^{\mu\nu\rho} A_{\mu} \partial_{\nu} A_{\rho},
\end{equation}
where for now $A_\mu$ represents an emergent gauge field that arises from the collective behavior of the electrons in the system as well as their coupling to a background electromagnetic field where the dynamical degrees of freedom have been integrated out,
and $\epsilon^{\mu\nu\rho}$ is the Levi-Civita symbol in three dimensions. This is a topological quantum field theory that describes the low energy effective field theory of the quantum Hall state and encodes the Hall conductivity $\sigma_{ij} = \frac{ \nu}{2 \pi}$ through $J_i = \frac{\delta S}{\delta A_{i}} = \frac{ \nu}{2 \pi} \epsilon_{ij}E_i $, where this is the expression for current $J_i$ that arises from the topological Chern-Simons action \cite{Tong}.

The Hall shift and viscosity terms can be described by adding the following Wen-Zee term to the effective action:
\begin{equation}
    S_{WZ}[A, \omega] = \frac{\nu \mathcal{S}}{4 \pi} \int d^3x \epsilon^{\mu\nu\rho} \omega_{\mu} \partial_{\nu} A_{\rho}.
\end{equation}
This is a mixed Chern-Simons term, where $\omega_\mu$ is the $SO(2)$-spin connection for local spatial rotations.
This Wen-Zee term encodes both the Hall shift $\mathcal{S}$ and viscosity $\eta_H = \nu \mathcal{S} B / 8 \pi$, which was derived in \cite{ReadRezayi}.

While these topological effective actions are enormously successful at describing many aspects of quantum Hall systems, they have some limitations. Consider, as Son does in \cite{DamSon2}, a microscopic theory corresponding to the initial topological Chern-Simons theory. This will be a theory of a system of non-relativistic electrons $\psi$
coupled to an external electromagnetic field $A_\mu$ and metric $h_{ij}$,
\begin{equation}
S=\int d^{3} x \sqrt{h}\left[\frac{i}{2} \psi^{\dagger} \stackrel{\leftrightarrow}{D}_{t} \psi-\frac{h^{i j}}{2 m} D_{i} \psi^{\dagger} D_{j} \psi\right],
\end{equation}
where $h_{ij}$ is a spatial metric, time is absolute, 
and $D_{\mu} = \partial_{\mu} - i A_{\mu}$ is the covariant derivative. This theory is invariant under the following coordinate reparametrizations: 
\begin{align} \label{gauge1}
    \delta \psi &= -\xi^k \partial_k \psi, \\
 \label{gauge1}
    \delta A_{0} &= -\xi^k \partial_k A_0 - A_k\dot{\xi}^k, \\
 \label{gauge2}
    \delta A_{i} &= -\xi^k \partial_k A_i - A_k\partial_i \xi^k - m(h_{ik}\dot\xi^k), \\
 \label{htrans}
    \delta h_{ij} &= -\xi^k \partial_k h_{ij} - h_{ij}\partial_i\xi^k - h_{ik}\partial_j\xi^k.
\end{align}
Here, these transformations of the fields correspond to introducing a diffeomorphism $x^\mu \rightarrow x^\mu + \xi^\mu (x,t)$. These transformations can be thought of as the non-relativistic limit of general coordinate invariance \cite{DamSon3}.

Considering that this topological theory describes a system of non-relativistic electrons, it should likewise be invariant under such transformations. However, under these same transformations, the Chern-Simons action \eqref{CS action} transforms as 
\begin{equation}
\delta S_{\mathrm{CS}}[A]=\frac{\nu}{2 \pi} \int d^{3} x \varepsilon^{i j}\left(m E_{i}\right) h_{j k} \dot{\xi}^{k}.
\end{equation}
This indicates that we need to find a better effective field theory description for such quantum Hall systems. Indeed, there are at least four reasons why we are interested in finding a more satisfactory effective field theory: (i) as mentioned, the standard Chern-Simons theory for the QHE should be invariant under coordinate transformations, but does not possess these standard non-relativistic symmetries;
(ii) when one considers further interaction terms for both microscopic and topological theories such as those including the electron $g$-factor, it turns out that this failure of coordinate invariance is even more dramatic and the topological theory can only account for situations in which $m \rightarrow 0$ and $g \rightarrow 0$; (iii) Son demonstrates that the topological theory doesn't behave correctly in the appropriate limits---so, clearly, the prior Chern-Simons theory cannot be the complete effective action for generic $m$ and $g$ (see \cite[\S2]{DamSon1} for further discussion on this point); (iv) a more general effective field theory for the QHE must also incorporate the additional Wen-Zee term to account for the Hall shift and viscosity while also respecting the above symmetry considerations.




Newton-Cartan theory then enters the picture in the following way. A Newton-Cartan geometry is a structure ($M$, $n_{\mu}$, $h^{\mu\nu}$, $\nabla$) where $M$ is a smooth manifold,  $n_{\mu}$ and $h^{\mu\nu}$ define a degenerate metric structure with $n_{\mu} n_{\nu}$ being the temporal metric (asssuming temporal orientability) and possessing a signature (1, 0, 0, 0) and $h^{\mu\nu}$ being the spatial metric and possessing a signature (0, 1, 1, 1), and $\nabla$ is a derivative operator \cite[ch.~4]{Malament}. Both metrics are compatible with the derivative operator and orthogonal to each other, such that
\begin{equation} \label{NCRelations}
\begin{split}
\nabla_{\lambda}n_{\mu} = 0 \,,
\qquad
\nabla_{\lambda}h^{\mu\nu} = 0 \,,
\qquad
n_{\mu}h^{\mu\nu} = 0 \,.
\end{split}
\end{equation}
Furthermore, one can also define a velocity vector field $v^\mu$ that satisfies $v^{\mu}n_{\mu} = 1$. One can also uniquely define a global time function $t$ such that $n_\mu = {(dt)}_\mu$ and this fixes the temporal metric while still leaving freedom to transform the spatial metric.
One of Son's key observations is that the spatial metric of this Newton-Cartan geometry $h^{\mu\nu}$ transforms under spatial diffeomorphisms $x^\mu \rightarrow x^\mu + \xi^\mu (x)$ in the same way that the metric transforms under transformations that correspond to the non-relativistic limit of general coordinate transformations in \eqref{htrans}. Additionally, the velocity vector $v^\mu$ transforms in the following way:
\begin{equation} \label{vtrans}
\delta v^{i} = -\xi^k \partial_k v^i + v^k \partial_k\xi^i +\dot{\xi}^i \,.
\end{equation}

The fact that the metric transforms in this way suggests that a Newton-Cartan background is the most natural framework for modelling a system respecting these kinds of symmetries. Furthermore, Son asks us to consider the following objects, which he calls the `improved gauge potentials':
\begin{equation}
\label{eq:x}
\begin{split}
\Tilde{A}_0 = A_0 - \frac{m}{2} v^2 - \frac{g}{4}e^{ij}\partial_i v_j\,,
\qquad
\Tilde{A}_i = A_i + m v_i\,.
\end{split}
\end{equation}
Using \eqref{gauge1}, \eqref{gauge2} and \eqref{vtrans}, one finds that these improved gauge potentials transform as a one form
and we can use these gauge potentials to define an effective quantum Hall action that possesses the required symmetries \cite{DamSon1}. That is, the improved gauge potentials now properly transform in such a way that they are invariant under the same transformations as those under which the microscopic theory is invariant. Son then goes on to write an effective action and demonstrates that this action behaves appropriately under salient limits and that one can compute additional features such as the Hall viscosity and corrections to the conductivity from the electromagnetic response. 

\subsection{Edge modes from Son's bulk model}\label{s2.2}

For our purposes, it will be most illuminating to look at the work in \cite{Hoyos} undertaken by Moroz \emph{et al.},~as these authors use Son's construction to examine the edge states of a quantum Hall system respecting the above symmetries (cf.~\cite{Gromov} for further discussion of edge states of the Newton-Cartan QHE). Their action is
\begin{equation} \label{Hoyos action}
\tilde{S}_{Q H}=-\frac{1}{4 \pi} \int d^{3} x \epsilon^{\mu \nu \lambda}\left(\frac{1}{\nu} a_{\mu} \partial_{\nu} a_{\lambda}+2 a_{\mu} \partial_{\nu} \tilde{\mathcal{A}}_{\lambda}\right),
\end{equation}
where $a_\mu$ now corresponds to the emergent gauge field referenced in Son's model and $\tilde{\mathcal{A}}_{\mu} := \tilde{A}_{\mu} + \mathcal{S} \tilde{\omega}_\mu$. Here $\tilde{A}_\mu$ is equivalent to Son's improved gauge potentials that transform correctly under non-relativistic spatial diffeomorphisms, and $\mathcal{S}\tilde{\omega}_\mu$ comes from the Wen-Zee term that encodes the Hall shift and includes the shift $\mathcal{S}$ and spin connection $\omega_\mu$. $\tilde{\omega}_\mu$ symbolizes that the spin connection has also been modified to transform properly under non-relativistic spatial diffeomorphisms, and is given by
\begin{equation} \label{wtrans}
\begin{split}
\tilde{\omega}_t = \omega_t +\frac{1}{2}\epsilon^{ij}\partial_i (h_{jk}v^k) \,,
\qquad
\tilde{\omega}_i = \omega_i \,.
\end{split}
\end{equation}

The action \eqref{Hoyos action} corresponds to \eqref{CS action}, save for the fact that the external gauge potentials have been modified to manifest the desired transformation properties and we have included an additional Wen-Zee term. In order to examine what happens on a boundary and explore the physics of edge modes, we first note that under a $U(1)$ gauge transformation $a_{\mu} \rightarrow a_{\mu} + \partial_{\mu} \lambda$ the action \eqref{Hoyos action} is invariant up to a total derivative term. Normally we can throw away total derivative terms, but in the presence of a boundary this term will not necessarily vanish. This is concerning because it seems to indicate that our theory is not gauge invariant in the presence of a boundary. One way to solve this problem---and the way this problem has been traditionally approached in the original papers on the subject such as \cite{Wen}---is to restrict the gauge choices in such a way to kill the boundary term and ensure invariance under gauge transformations. For instance, under the above gauge transformation (and identifying $\partial_{\mu} \lambda = \delta a_{\mu}$),
the action \eqref{Hoyos action} changes by a term proportional to $\delta S_{M} = \int d^3x \epsilon^{\mu\nu\rho} \partial_{\mu} (a_{\nu} \delta a_{\rho})$. In the presence of a boundary at $y=0$ along the $x$-direction, we can use Stokes' theorem to turn this into a surface integral $\delta S_{\partial M} = \int dx dt (a_t \delta a_x - a_x \delta a_t)$. Using the gauge fixing condition $a_t + v a_x = 0 $ (where $v$ is a parameter with units of velocity), as the authors of \cite{Hoyos} do, clearly kills this boundary term. One then plugs this gauge fixing condition back into the bulk action \eqref{Hoyos action}, examines what happens at the boundary $y = 0$, and finds the action for a chiral boson on the boundary (see \cite{Tong}, [p.205]).
The traditional interpretation, as stated by Wen and often repeated in the physics literature, is that ``the Chern-Simons action is not invariant under gauge transformations ... due to the boundary effects. To solve this problem we will restrict the gauge transformations to be zero on the boundary. ... Due to this restriction some degrees of freedom of $a_{\mu}$ on the boundary become dynamical'' \cite{Wen}. In essence, this restriction of gauge choice is understood to manifest itself as the resulting dynamics on the boundary. 

Moroz \emph{et al.}~follow a similar procedure and substitute this gauge fixing condition into their bulk action \eqref{Hoyos action} and use it to eliminate the temporal components. They then use the equations of motion for the gauge field as well as Gauss's law to write the gauge field as $a_i = \partial_i \theta - \nu \tilde{\mathcal{A}}_i =: \tilde{D}_i \theta$, integrate by parts using Stokes' theorem as before, and find the boundary action for a chiral boson that lives on $y=0$: 
\begin{equation} \label{HoyosMoroz}
S_{\theta}=\frac{1}{4 \pi} \int d^{2} x\left[\frac{1}{\nu}\left(\tilde{D}_{t} \theta+v^{x} \tilde{D}_{x} \theta\right) \tilde{D}_{x} \theta-\theta \tilde{\mathcal{E}}_{x}\right].
\end{equation}
Here $\theta$ is the chiral boson and $\tilde{\mathcal{E}}_x = \tilde{E}_x + s \tilde{E}_{\omega x} $, where $\tilde{E}_x = \partial_t \tilde{A}_x - \partial_x \tilde{A}_t$ and $\tilde{E}_{\omega x} = \partial_t \tilde{\omega}_x - \partial_x \tilde{\omega}_t$.

This gives the action for the chiral boson at the boundary of a quantum Hall system that is invariant under non-relativistic diffeomorphisms and also encodes additional phenomena such as the Hall shift and viscosity \cite{Hoyos}.\footnote{In calculating certain quantities from this action such as the longitudinal electric conductivity, as undertaken by the authors of \cite{Hoyos}, it was pointed out in \cite{MorozReply, MorozCorrection} that one must pay careful attention to whether the velocity field is a dependent or independent quantity.} While certainly useful in describing the physics of quantum Hall systems, this was all derived in the language of non-relativistic diffeomorphism invariance and not in the language of explicit Newton-Cartan objects. That is, the action here respects the same transformation rules for diffeomorphisms as the background Newton-Cartan geometry does, but the actual roles of the Newton-Cartan objects are obscured by this language. Furthermore, as we shall see, this formalism has an important limitation.

\section{Milne invariant effective action}\label{s3}

In this section, we demonstrate how Son's bulk model can be reformulated using an approach more natural to the Newton-Cartan structures based on a Milne invariant effective action. 

Jensen points out that Son's non-relativistic diffeomorphism invariance can actually be understood as special cases of a Milne boost \cite{Jensen1}. Milne boosts are a symmetry of the Newton-Cartan structure and when Newton-Cartan theory is cast in the vielbein formalism, one sees that Milne boosts are nothing other than local Galilean boosts \cite{GeraciePrabhu}. To explain further, first note that the velocity vector $v^\mu$ introduced when we discussed Newton-Cartan geometry satisfies $v^{\mu}n_{\mu} = 1$ and is ambiguous up to $(v^{\mu} + h^{\mu\nu}\psi_{\nu}) n_{\mu} = 1 $, where $\psi_\mu$ is a spatial one form such that $v^{\mu}\psi_{\mu} = 0$. In order to preserve the defining Newton-Cartan relationships \eqref{NCRelations}, we demand that these Newton-Cartan objects transform in the following way and these define the Milne boosts:
\begin{equation} \label{Milne}
\begin{aligned}
n_{\mu} \rightarrow n_{\mu}^{\prime} &=n_{\mu} , \quad h^{\mu \nu} \rightarrow h^{\prime \mu \nu}=h^{\mu \nu}, \\
v^{\mu} \rightarrow v^{\prime \mu} &=v^{\mu}+h^{\mu \nu} \psi_{\nu}, \\
h_{\mu \nu} \rightarrow h_{\mu \nu}^{\prime} &=h_{\mu \nu}-\left(n_{\mu} P_{\nu}^{\rho}+n_{\nu} P_{\mu}^{\rho}\right) \psi_{\rho}+n_{\mu} n_{\nu} h^{\rho \sigma} \psi_{\rho} \psi_{\sigma}.
\end{aligned}
\end{equation}
Notice also that $h^{\mu\nu}$ does not transform the same way as its inverse $h_{\mu\nu}$, where the latter transformation involves the quantity $P_{\mu}^{\rho} := h_{\mu \nu} h^{\nu \rho}=\delta_{\mu}^{\rho}-n_{\mu} v^{\rho}$. The transformations taking this form can be traced to the degenerate metric structure. Alternatively, these transformations can be equivalently formulated such that $(v^{\mu}, h_{\mu\nu})$ are invariant while $(n_{\mu}, h^{\mu\nu})$ instead transform under the Milne boosts \cite{Ban1}.

There is another piece to this story. For Newton-Cartan geometry (and unlike Lorentzian geometry) the metric compatibility conditions only pick out a space of connections. A particular connection within this class is then by further specifying a two-form $F$. That is, 
\begin{equation}
\begin{aligned}
 &\Gamma^{\mu}{ }_{\nu \rho}=v^{\mu} \partial_{\rho} n_{v}+\frac{1}{2} h^{\mu \sigma}\left(\partial_{v} h_{\rho \sigma}+\partial_{\rho} h_{v \sigma}-\partial_{\sigma} h_{v \rho}\right)+h^{\mu \sigma} n_{(v} F_{\rho), \sigma} 
\end{aligned}
\end{equation}
where we identify this two-form as the field strength corresponding to what is called the `mass gauge field' $a_\mu$. In order to preserve invariance under Milne boosts for the connection, there is a unique way that the mass gauge field must transform. This is given by
\begin{equation}\label{gaugefieldboost}
    a_{\mu}' = a_{\mu} +  \psi_{\mu} -\frac{1}{2} n_{\mu} \psi^2.
\end{equation}
For this reason, the mass gauge field is often taken to be part of the Newton-Cartan data along with $(M, n_{\mu}, h^{\mu\nu}, \nabla )$, and later on we will associate this mass gauge field with the emergent gauge field seen in the QHE.

We can now better understand why there is something unsatisfactory about using the improved gauge potentials constructed from non-relativistic diffeomorphism transformations. Defining $\tilde{A}_{\mu}$ as Son does effectively picks out a special coordinate system because Son's non-relativistic diffeomorphism invariance is a special case of Milne boosts where the velocity parameter is held under particular constraints (see \cite[\S2.7]{Jensen1} for a discussion on the relationship between Milne boosts and non-relativistic coordinate invariance).
This is problematic because it is unnatural to demand that these background electromagnetic fields transform in a way which picks out a specific coordinate system and we seek a construction that better clarifies the role of this symmetry of the Newton-Cartan geometry.

Geracie {\em et al.}~tackle in \cite{GeraciePrabhu} the same problems as Son, but take up the above challenge. Rather than imposing diffeomorphism invariance by hand as Son does with his modified diffeomorphisms, which are subject to the aforementioned constraints, they `dress' the gauge field to be invariant under Milne boosts. (Note that Geracie {\em et al.}~do not explicitly identify their approach here as one case of the general `dressing field' methodology described in e.g.~\cite{Francois}; we discuss this methodology further below.) In order to do this, they define an effective gauge field,
\begin{equation}
    \mathcal{A}_{\mu} := A_{\mu} + m \overset{I}{a}_{\mu},
\end{equation}
where $A_{\mu}$ is the background electromagnetic field and $m \overset{I}{a}_{\mu}$ is the dressed mass gauge field from the Newton-Cartan construction. What does this `dressing' amount to? As we have seen above with (\ref{gaugefieldboost}), the mass gauge field $a_{\mu}$ is sensitive to Milne boosts and is not invariant under these transformations. To address this, they dress the mass gauge field to be invariant under Milne boosts and this dressed up mass gauge field will be denoted $\overset{I}{a}_\mu$. How is this achieved? Consider a Milne invariant vector field $u^{\mu}$ that is timelike and normalized such that $n_{\mu}u^{\mu}=1$. We can lower the index using $u_{\mu} = h_{\mu\nu}u^{\nu}$, and this lowered $u_{\mu}$ will inherit its Milne transformation properties from $h_{\mu\nu}$ in \eqref{Milne}. One finds that we can find a combination of terms using this object that transforms in the exact opposite way as the mass gauge field $a_{\mu}$. In particular, one finds that the following object transforms under Milne boosts in this way \cite{Jensen2}:
\begin{equation}
    \left(u_{\mu} - \frac{1}{2}u^2 n_{\mu}\right)' = u_{\mu} - \frac{1}{2}u^2 n_{\mu} -\psi_{\mu} + \frac{1}{2} n_{\mu} \psi^2.
\end{equation}
We then use this object to dress the mass gauge field, which is given by
\begin{equation}
    m \overset{I}{a}_{\mu} = m \left( a_{\mu} + u_{\mu} - \frac{1}{2} u^2 n_{\mu} \right).
\end{equation}
One can easily see that this dressed mass gauge field is invariant under Milne boosts as the transformations of $a_{\mu}$ and $u_{\mu}$ exactly cancel out each other.

Now that we have a Milne invariant mass gauge field, we can construct an effective quantum Hall action given by
\begin{equation} \label{GPAction}
    L = \frac{\nu}{4\pi} \epsilon^{\mu\nu\rho} \mathcal{A}_{\mu} \partial_{\nu} \mathcal{A}_{\rho} + \frac{k}{2\pi} \epsilon^{\mu\nu\rho} \omega_{\mu} \partial_{\nu} \mathcal{A}_{\rho},\footnote{Geracie {\em et al.}~include more terms in their action because they are interested in some non-topological effects. As we are only interested in the edge modes derived from the topological field theory, we retain only the topological terms.}
\end{equation}
where $\mathcal{A}_\mu$ is our effective gauge field that is Milne invariant due to how $\overset{I}{a}_{\mu}$ has been dressed for Milne invariance and $\omega_\mu$ is the spin connection \cite{GeraciePrabhu}. We see that the first term is the Chern-Simons term and the second term is the Wen-Zee term. This action corresponds to \eqref{Hoyos action}, but rather than having imposed diffeomorphism invariance on the background electromagnetic fields, we have dressed the mass gauge field to be Milne invariant. This has the conceptual advantage of separating out background electromagnetic fields defined only in a particular coordinate system from the local Milne boost invariance that a system of non-relativistic particles will have. That is, the emergent gauge field that arises from the collective, non-relativistic behavior of the electrons is now identified with a Milne invariant mass gauge field from the Newton-Cartan structure and the background electromagnetic fields behave as normal. This action is now formulated explicitly in terms of Newton-Cartan objects and we will now apply the Donnelly-Freidel programme to look at the edge modes of this system.

\section{Newton-Cartan edge modes}\label{s4}

In this section, we first recall the essential details of the Donnelly-Freidel approach to gauge symmetries in systems with boundaries (\S\ref{s4.1}); then, we apply this machinery to the above-described quantum Hall effective action of Geracie \emph{et al.}~(\S\ref{s4.2}).

\subsection{The Donnelly-Freidel programme}\label{s4.1}

The Donnelly-Freidel programme in essence proposes that one must enhance the original gauge symmetries on the boundary of systems in order to truly capture the empirical content of gauge symmetries \cite{DL}. Recall that the traditional way of understanding edge modes in the quantum Hall effect relies on restricting the gauge transformations in such a way that the theory is gauge invariant in the presence of a boundary,
and physicists have traditionally interpreted these constraints as sourcing the dynamics of these edge modes. On the contrary, the Donnelly-Freidel programme shows that we can understand edge modes as resulting from the need to enhance the gauge symmetries on the boundary; i.e.,~our theory is incomplete and the failure of gauge invariance is telling us that we are missing something from the theory. The Donnelly-Freidel programme then offers a systematic way of identifying what we are missing, completing the theory, and describing edge modes in the presence of a boundary without the need to ever place constraints on allowed gauge transformations. 

Donnelly and Freidel make extensive use of the covariant Hamiltonian formalism. In briefly reviewing this formalism, we follow the formal manipulations of \cite{Geiller}, but direct the reader to \cite{LMP} for a more precise construction. Consider a Lagrangian $L [\Phi]$ that is a function of fields $\Phi$ and a potential boundary term d$b[\Phi]$. The variation of the Lagrangian is
\begin{equation}\label{eq:4.1}
    \delta L_b [\Phi] = \delta (L[\Phi] + db[\Phi]) =  E[\Phi] \wedge \delta \Phi + d\theta_{b,c} [\Phi, \delta \Phi].
\end{equation}
Here the first term corresponds to the equations of motion and vanishes on-shell. The second term is the `pre-symplectic potential'. We can write $\theta_{b,c}$ as
\begin{equation}
    \theta_{b,c}[\Phi, \delta \Phi] = \theta [\Phi, \delta \Phi] + \delta b [\Phi] + dc [\Phi, \delta \Phi].
\end{equation}
This object is a $(d-1)$-form in spacetime and a one-form in field space. We also notice that due to the exterior derivative in front of $\theta_{b,c}$, its identification is ambiguous up to a corner term $c$.\footnote{Later, we will fix the corner term by adding a boundary action as in \cite{LMP}.} We then define the `pre-symplectic form' as follows:
\begin{equation}
    \omega_{b,c}[\Phi, \delta_1, \delta_2] = \delta \theta_{b,c} [\Phi, \delta \Phi].
\end{equation}
Here $\delta$ corresponds to a field space derivative. The pre-symplectic form is a $(d-1)$-form in spacetime and a two-form in field space.

These objects will be crucial for us in investigating the behavior on the boundary of our quantum Hall system living in a Newton-Cartan spacetime. In anticipation, we note that we will generally find that our pre-symplectic potential $\theta_{b,c} [\Phi, \delta \Phi]$ will not be invariant under a gauge transformation $\Phi \rightarrow \Phi + d\alpha$. That is, $\theta_{b,c} [\Phi + d\alpha, \delta(\Phi + d\alpha)] \neq \theta_{b,c} [\Phi, \delta \Phi]$. The key insight learned from this fact is that our pre-symplectic potential is incomplete and this suggests that there is an important piece of our physics missing. This suggests that we must introduce a new field $u$ on the boundary that transforms under the gauge transformation as $u \rightarrow u -\alpha$ in order to successfully restore gauge invariance on the boundary. That is, $\theta_{b,c} [\Phi + d\alpha, \delta (\Phi + d\alpha), u - \alpha, \delta (u - \alpha)] =  \theta_{b,c} [\Phi, \delta \Phi, u, \delta u]$, and we have enhanced the gauge symmetry of our original system. We then can derive the pre-symplectic form and explore what this new field does on the boundary. The pre-symplectic form $\omega_{b,c}$ can be integrated over field space such that  $\Omega_{\Sigma,\partial \Sigma} = \int_{\Sigma} \omega [\Phi, \delta \Phi] + \int_{\partial \Sigma} \omega [\Phi, \delta \Phi, u, \delta u]$, where $\Sigma$ is a ($d-1$)-dimensional hypersurface and the term on the boundary gives the resulting pre-symplectic form in terms of this new field that restores gauge symmetry and exists only on the boundary. 

\subsection{Donnelly-Freidel and the Newton-Cartan quantum Hall action}\label{s4.2}

Recalling that varying the Lagrangian takes the form \eqref{eq:4.1},
we will now be interested in looking at the boundary observables of a quantum Hall system in a Newton-Cartan background geometry. Following the Donnelly-Freidel programme, we will see how boundary observables for this system arise from demanding gauge invariance of the pre-symplectic potential. 

The general action for a quantum Hall system in a Newton-Cartan geometry, given by the Lagrangian in \eqref{GPAction}, can be written equivalently in the notation of differential forms as
\begin{equation}\label{formaction}
    S = \int_{M} \mathcal{A} \wedge d\mathcal{A} + 2 \int_{M} \omega \wedge d\mathcal{A} =: S_{CS}+S_{WZ},
\end{equation}
where, again, $\mathcal{A} = A + m\overset{I}{a}$ is the improved gauge field that is dressed to be Milne invariant.
Varying the Lagragians associated with both pieces of this action, we have
\begin{equation}\label{CSvariation}
\begin{aligned}
    \delta L_{CS}[\mathcal{A}] &= \delta \mathcal{A} \wedge d\mathcal{A} + \mathcal{A} \wedge  d\delta\mathcal{A} \\ &= \delta \mathcal{A} \wedge d\mathcal{A} + d(\mathcal{A} \wedge  \delta\mathcal{A}) - d\mathcal{A} \wedge \delta \mathcal{A}  \\ &= 2\delta \mathcal{A} \wedge d\mathcal{A} + d(\delta\mathcal{A} \wedge  \mathcal{A}),
\end{aligned}
\end{equation}
\begin{equation}
\begin{aligned}
    \delta L_{WZ} [\omega, \mathcal{A}] &=  \omega \wedge d (\delta \mathcal{A}) \\ &= - d \omega \wedge \delta \mathcal{A} + d (\omega \wedge \delta \mathcal{A}).
\end{aligned}
\end{equation}
We identify the pre-symplectic potential from the action \eqref{formaction} to be $d\Theta = d\theta_{CS} + d\theta_{WZ} =  d(\delta\mathcal{A} \wedge \mathcal{A}) + d(\omega \wedge \delta \mathcal{A}) $. This action is still Milne invariant by virtue of the dressed gauge field $\overset{I}{a}$; however, there is still the residual gauge freedom to make a $U(1)$ transformation. Recall also that this pre-symplectic potential will not be invariant under such a gauge transformation. We want \textit{both} Milne invariance and $U(1)$ invariance; thus, we will now compute how this pre-symplectic potential transforms under this gauge transformation in order to demonstrate how we can restore gauge invariance by introducing an additional field on the boundary along the lines suggested by Donnelly and Freidel.

\subsubsection{Chern-Simons term}\label{s4.3}
We begin with the Chern-Simons term of eq.~(\ref{CSvariation}) and expand out the terms from the effective gauge field:
\begin{equation}
\begin{aligned}
    \theta_{CS} \left[A, \overset{I}{a}, \delta \overset{I}{a}\right] &=
    \delta\mathcal{A}  \wedge \mathcal{A}  \\ &= \delta (A + m \overset{I}{a}) \wedge (A + m \overset{I}{a}) = m(\delta\overset{I}{a} \wedge A) +  m^2(\delta\overset{I}{a} \wedge \overset{I}{a}).
\end{aligned}
\end{equation}
Considering a gauge transformation such that  $\overset{I}{a} \rightarrow \overset{I}{a} + d\alpha$ gives
\begin{equation}
\begin{aligned}
     \theta_{CS} \left[A, \overset{I}{a} + d\alpha, \delta (\overset{I}{a} + d\alpha)\right] &=  m(\delta(\overset{I}{a}+d\alpha) \wedge A) + m^2(\delta(\overset{I}{a}+d\alpha) \wedge (\overset{I}{a}+d\alpha)).
\end{aligned}
\end{equation}
We now break this equation up into the parts that comprise the original pre-symplectic potential and the parts coming from the gauge transformation:
\begin{equation}
\begin{aligned}
\theta_{CS} \left[A, \overset{I}{a} + d\alpha, \delta (\overset{I}{a} + d\alpha)\right] & = m(\delta\overset{I}{a} \wedge A) +  m^2(\delta\overset{I}{a} \wedge \overset{I}{a}) \\ & \qquad +m^2\left[\delta \overset{I}{a} \wedge \mathrm{d} \alpha+\delta (\mathrm{d} \alpha) \wedge(\overset{I}{a}+\mathrm{d} \alpha)\right] +  m\left[\delta (d \alpha) \wedge A\right] \\
& =:  \theta_{CS} \left[A, \overset{I}{a}, \delta \overset{I}{a}\right] + \theta_1 \left[\overset{I}{a}, \delta \overset{I}{a}, \alpha, \delta \alpha \right] + \theta_2 \left[A, \delta \alpha \right] 
\end{aligned}
\end{equation}

\begin{equation}
\begin{aligned}
\theta_1 \left[\overset{I}{a}, \delta \overset{I}{a}, \alpha, \delta \alpha \right] &=m^2\left[\delta \overset{I}{a} \wedge \mathrm{d} \alpha+\delta (\mathrm{d} \alpha) \wedge(\overset{I}{a}+\mathrm{d} \alpha)\right] \\
&=m^2\left[\delta \overset{I}{a} \wedge \mathrm{d} \alpha+\mathrm{d}(\delta \alpha \wedge (\overset{I}{a}+\mathrm{d} \alpha))-\delta \alpha \wedge \mathrm{d} \overset{I}{a}\right] \\
&=m^2\left[\delta(\overset{I}{a} \wedge \mathrm{d} \alpha)-\overset{I}{a} \wedge \delta \mathrm{d} \alpha+\mathrm{d}(\delta \alpha \wedge (\overset{I}{a}+\mathrm{d} \alpha))-\delta \alpha  \wedge \mathrm{d} \overset{I}{a}\right] \\
&=m^2\left[\delta(\overset{I}{a} \wedge \mathrm{d} \alpha)-\delta \alpha \wedge \mathrm{d} \overset{I}{a}+\mathrm{d}(\delta \alpha \wedge \overset{I}{a})+\mathrm{d}(\delta \alpha \wedge (\overset{I}{a}+\mathrm{d} \alpha))-\delta \alpha \wedge \mathrm{d} \overset{I}{a}\right] \\
&=m^2\left[\mathrm{d}(\delta \alpha \wedge (2 \overset{I}{a}+\mathrm{d} \alpha))+\delta(\overset{I}{a} \wedge \mathrm{d} \alpha)-2 \delta \alpha \wedge \mathrm{d} \overset{I}{a}\right]
\end{aligned}
\end{equation}

\begin{equation}
\begin{aligned}
\theta_2 \left[A, \delta \alpha \right] &=m\left[ \delta (d \alpha) \wedge A\right] = m\left[ - \delta\alpha \wedge dA + d( \delta \alpha \wedge A)\right] 
\end{aligned}
\end{equation}

We can thereby see that the gauge-transformed part of the pre-symplectic potential originating from the Chern-Simons term can be written as
\begin{equation}
\begin{aligned}
\theta_{CS} \left[A, \overset{I}{a} + d\alpha, \delta (\overset{I}{a} + d\alpha)\right] &= \theta_{CS} \left[A, \overset{I}{a}, \delta \overset{I}{a}\right] + \theta_1 \left[\overset{I}{a}, \delta \overset{I}{a}, \alpha, \delta \alpha \right] + \theta_2 \left[A, \alpha, \delta \alpha \right] \\
&=  m(\delta\overset{I}{a} \wedge A) +  m^2(\delta\overset{I}{a} \wedge \overset{I}{a})  + m^2\left[\mathrm{d}(\delta \alpha \wedge (2 \overset{I}{a}+\mathrm{d} \alpha))+\delta(\overset{I}{a} \wedge \mathrm{d} \alpha)-2 \delta \alpha \wedge\mathrm{d} \overset{I}{a}\right] \\
&\qquad + m\left[ - \delta\alpha \wedge dA + d( \delta \alpha \wedge A)\right].
\end{aligned}
\end{equation}

\subsubsection{Wen-Zee term}\label{s4.4}
We now do the same for the Wen-Zee term:
\begin{equation}
    \theta_{WZ} \left[\omega, \delta \overset{I}{a}\right] = \omega \wedge \delta \mathcal{A}  = \omega \wedge \delta (A + m \overset{I}{a}) = m \left[\omega \wedge \delta \overset{I}{a}\right].
\end{equation}
We likewise consider the gauge transformation  $\overset{I}{a} \rightarrow \overset{I}{a} + d\alpha$ and write the gauge transformed Wen-Zee term as,
\begin{equation}
\begin{aligned}
\theta_{WZ} \left[\omega, \delta (\overset{I}{a} + d\alpha)\right] &=  m \left[\omega \wedge \delta (\overset{I}{a}+ d\alpha)\right] = m \left[\omega \wedge \delta \overset{I}{a}\right] + m \left[\omega \wedge \delta (d\alpha)\right] \\
&=: \theta_{WZ} \left[\omega, \delta A, \delta \overset{I}{a}\right] + \theta_3\left[\omega, \delta \alpha \right],
\end{aligned}
\end{equation}
where
\begin{equation}
    \theta_3\left[\omega, \delta \alpha \right] = m \left[\omega \wedge \delta (d\alpha)\right] = m\left[d (\omega \wedge \delta \alpha) - d\omega \wedge \delta \alpha\right] .
\end{equation}

\subsubsection{Gauge-transformed pre-symplectic potential}\label{s4.5}

We can finally write the gauge-transformed pre-symplectic potential as
\begin{equation}
\begin{aligned}
\Theta &= \theta_{CS} \left[A, \overset{I}{a}, \delta \overset{I}{a}\right] + \theta_1 \left[\overset{I}{a}, \delta \overset{I}{a}, \alpha, \delta \alpha \right] + \theta_2 \left[A, \delta \alpha \right] + \theta_{WZ} \left[\omega, \delta \overset{I}{a}\right] + \theta_3\left[\omega, \delta \alpha \right] \\
&= m(\delta\overset{I}{a} \wedge A) +  m^2(\delta\overset{I}{a} \wedge \overset{I}{a}) +m^2\left[\mathrm{d}(\delta \alpha \wedge (2 \overset{I}{a}+\mathrm{d} \alpha))+\delta(\overset{I}{a} \wedge \mathrm{d} \alpha)-2 \delta \alpha \wedge \mathrm{d} \overset{I}{a}\right] \\
&\qquad+m\left[ - \delta\alpha \wedge dA + d( \delta \alpha \wedge A)\right] +  m \left[\omega \wedge \delta \overset{I}{a}\right] +  m\left[d (\omega \wedge \delta \alpha) - d\omega \wedge \delta \alpha\right]
\end{aligned}
\end{equation}



As noted before, this is clearly not invariant under a gauge transformation, but we can restore gauge invariance by introducing a new field on the boundary. This will be a dressing gauge field $u$ that transforms as $u\rightarrow u -\alpha $ and effectively dresses the pre-symplectic potential to be invariant under $U(1)$ gauge transformations. This is analogous to what we did before in dressing the mass gauge field to be Milne invariant. In fact, by replacing every $\alpha$ with a $u$ in the pre-symplectic potential, we then find that this new pre-symplectic potential is gauge-invariant. Let us write down this new pre-symplectic potential $\tilde{\Theta}$:
\begin{equation}
\begin{aligned}
\tilde{\Theta} &= \theta_{CS} \left[A, \overset{I}{a}, \delta \overset{I}{a}\right] + \theta_{WZ} \left[\omega, \delta \overset{I}{a}\right] + \theta_1 \left[\overset{I}{a}, \delta \overset{I}{a}, u, \delta u \right] + \theta_2 \left[A, \delta u \right] + \theta_3\left[\omega, \delta u \right] \\
&= \theta_{CS} \left[A, \overset{I}{a}, \delta \overset{I}{a}\right] + \theta_{WZ} \left[\omega, \delta \overset{I}{a}\right] \\
&\qquad+m^2\left[\mathrm{d}(\delta u \wedge (2 \overset{I}{a}+\mathrm{d} u))+\delta(\overset{I}{a} \wedge \mathrm{d} u)-2 \delta u \wedge \mathrm{d} \overset{I}{a}\right] \\
&\qquad+m\left[ - \delta u \wedge dA + d( \delta u \wedge A)\right] +  m\left[d (\omega \wedge \delta u) - d\omega \wedge \delta u\right].
\end{aligned}
\end{equation}
This new pre-symplectic potential is now invariant under $U(1)$ gauge transformations
such that $\overset{I}{a} \rightarrow \overset{I}{a} + d\alpha$ because the new field $u$ transforms as $ u \rightarrow u -\alpha$, which kills all the additional terms that result from the transformation on the original pre-symplectic potential $\theta_{CS} + \theta_{WZ}$.\footnote{E.g., a term such as $m(\delta(\overset{I}{a}+d\alpha) \wedge A)$ would be replaced with $m(\delta(\overset{I}{a}+ du) \wedge A)$. This would now transform as $m(\delta(\overset{I}{a}+du) \wedge A)$ $\rightarrow$ $m(\delta(\overset{I}{a} + d\alpha + du - d\alpha) \wedge A) = m(\delta(\overset{I}{a}+ du) \wedge A)$. All the terms in the new pre-symplectic potential $\tilde{\Theta}$ transform in this way, which makes it now $U(1)$ gauge invariant.} That is, our addition of $u$ into $ \tilde{\Theta}$ has successfully restored gauge invariance under this $U(1)$ symmetry. 

As we anticipated before, the introduction of this new field has interesting and important consequences on the boundary. Recall that the pre-symplectic current is defined as $ \omega_{b,c} =  \delta \theta_{b,c}$  \cite[eq.~2.3]{Geiller}. Computing this object by taking $\omega_{b,c} = \delta \tilde{\Theta}$ and integrating $\Omega_{\Sigma,\partial \Sigma} = \int_{\Sigma} \omega + \int_{\partial \Sigma} \omega$,
we find the bulk and boundary contributions,
\begin{equation}\label{boundaryform}
\begin{aligned}
 \Omega_{\Sigma,\partial \Sigma} = \int_{\Sigma}  \delta \overset{I}{a} \wedge \delta \overset{I}{a } + \int_{\partial \Sigma} \delta u \wedge \delta (2\overset{I}{a}+du).
\end{aligned}
\end{equation}

This agrees with Geiller's derivation of the symplectic structure for standard Chern-Simons theory (ours being different in that the gauge field $a_{\mu}$ has been dressed to be Milne invariant and that the gauge field is coupled to two background fields representing the electromagnetic field and spin connection). Furthermore, we see that the background fields present in the pre-symplectic potential drop out of the pre-symplectic form due to the application of the field space derivative in calculating the pre-symplectic form. Notice also that the new field $u$ lives only on the boundary. Additionally, both \cite{DL} and \cite{Geiller} note that this construction leads to the emergence of a new symmetry on this boundary. That is, there is a symmetry that acts on the original fields as $\overset{I}{a} \rightarrow \overset{I}{a}$, $A \rightarrow A$, $\omega \rightarrow \omega$, and the boundary field as $u \rightarrow u + \mathcal{B}$, where $\mathcal{B}$ is a constant shift.
Furthermore, they note that this new boundary symmetry is actually generated by the new boundary field, which corresponds to the observable edge modes of the theory. We will discuss the empirical implications of this symmetry in the following section.

This result gives the pre-symplectic form for the boundary of the theory. However, as is noted in \cite{LMP,TR}, the symplectic structure is an output of a dynamical variational principle, yet the Donnelly-Freidel programme as originally construed does not tell us the action that produces the symplectic structure noted above that contains this new boundary field that restores the gauge invariance of the theory. In this spirit, we propose the following action:
\begin{equation}\label{NEWaction}
\begin{aligned}
    S = \int_{M} \mathcal{A} \wedge d\mathcal{A} + 2 \int_{M} \omega \wedge d\mathcal{A} + \int_{\partial M}  du \wedge \overset{I}{a} + (du + \overset{I}{a}) \wedge * (du + \overset{I}{a}) +  du \wedge 2 (\omega + A).
\end{aligned}
\end{equation}
Here the bulk terms are just the action for the bulk Newton-Cartan QHE model in \eqref{formaction}, the boundary term is the proposed contribution from the new boundary field, and $*$ is the boundary Hodge star operator. This is similar to the proposal in \cite{LMP}, where they examine edge modes for standard Chern-Simons theory along the lines of the Donnelly-Freidel programme. 
If one follows the procedures for deriving the pre-symplectic form, the bulk terms of course give rise to the normal symplectic structure on $\Sigma$, whereas the boundary term reproduces the symplectic structure on $\partial \Sigma$ given in \eqref{boundaryform}. Furthermore, the boundary term restores the gauge invariance of the action as the terms with the new boundary field cancel the terms in the bulk when a gauge transformation is introduced, and this leaves us with a theory for the Newton-Cartan QHE that is both Milne and $U(1)$ gauge invariant. 

With this action in hand, we can readily study the dynamics of the resulting edge modes. Consider a variation of the boundary action: 
\begin{equation}\label{varyboundaryaction}
    \delta S_{\partial M} = \int_{\partial M} \delta (du) \wedge \left[ \overset{I}{a} +  * 2 (du + \overset{I}{a}) + 2(\omega +A) \right] + \delta (\overset{I}{a}) \wedge \left[ du + * 2 (du + \overset{I}{a}) \right].
\end{equation}
When the variation vanishes, we obtain $ \overset{I}{a} + * 2 (du + \overset{I}{a}) + 2(\omega +A) = 0$. Notice also that a variation of the field $\overset{I}{a}$ provides a constraint on the boundary such that $du + * 2 (du + \overset{I}{a}) = 0$. We can then use the second equation to re-write the first as $\overset{I}{a} -du + 2(\omega +A) = 0$. This is precisely the equation that Moroz \emph{et al.}~use in \cite{Hoyos} to write down the form of their emergent gauge field. We can then look at the action for the edge modes,
\begin{equation}\label{NEWboundaryaction}
    S_{\partial M} = \int_{\partial M}  du \wedge \overset{I}{a} + (du + \overset{I}{a}) \wedge * (du + \overset{I}{a}) +  du \wedge 2 (\omega + A)
\end{equation}
and when one identifies $u$ as the chiral boson, $\overset{I}{a}$ as the emergent gauge field, $\omega$ as the spin connection, and $A$ as a background electromagnetic field as these objects are understood in \cite{Hoyos}, and likewise uses the gauge fixing conditions found therein, we find that our proposal agrees with \eqref{HoyosMoroz}.





To summarise: in this section, we have applied the Donnelly-Freidel programme to a bulk model of the Newton-Cartan QHE. This model is a reformulation of Son's original bulk model of the Newton-Cartan QHE and was specifically formulated in terms of Milne invariant Newton-Cartan objects. In applying the Donnelly-Freidel programme to this model, we enhanced the symmetries of the model in the presence of a boundary by adding a new boundary field that restores the $U(1)$ gauge invariance of the theory without ever resorting to gauge fixing. This allowed us to construct the resulting edge modes of this Newton-Cartan QHE model by computing the pre-symplectic form, examining the boundary contributions, and building a new action that includes the boundary field and respects the symmetry requirements of the theory. 


\section{Empirical significance of symmetries}\label{s5}


As mentioned above, both the QHE and the existence of the resulting edge modes have been sources of puzzlement in the past. The original justification for the existence of edge modes was that we must restrict the allowed gauge transformations in such a way that the boundary term vanishes and gauge invariance is restored (recall, for example, the above-presented quote from Wen).
Then, the thinking goes, we can understand that these constraints on the allowed gauge transformations source the physical degrees of freedom that we see in the form of edge modes. However, having now followed the Donnelly-Freidel programme, we can see that it is not constraints on the gauge transformations which lead to the existence of edge modes, but rather that these are the result of the additional boundary fields that restore the $U(1)$ gauge invariance of the theory in the presence of the boundary.
Furthemore, we also note that the introduction of these boundary fields generates a new boundary symmetry that is only apparent after enhancing the original symmetries of the system by adding a boundary field to restore $U(1)$ gauge invariance of the model. With all of this in mind and considering that this Newton-Cartan model of the QHE required introducing a further symmetry related to the Milne invariance of the Newton-Cartan structure, the question we address in this section is this: how does this all fit into our understanding of the empirical significance of the relevant symmetries in this model? 

The traditional view on the empirical significance of symmetries distinguishes between global symmetries---such as rigid Galilean boosts---and local symmetries---such as the $U(1)$ symmetry in electromagnetism or in our present model. The traditional view holds that global symmetries---those symmetries which act identically at each spacetime point---have `direct empirical significance' (a definition of which will be offered below), whereas local symmetries---those which can act differently from spacetime point to spacetime point---do not. (For important introductions to this literature, see \cite{BB, GW}.)


Consider, for example, the well-known thought experiment of Galileo's ship. Imagine that you are in the interior of the ship while it is at rest, performing various experiments and noticing the relative positions and motions of objects in the interior with respect to you. Now imagine that the ship is moving at a constant velocity. Your experience in the interior will be indistinguishable from your experience when the ship was at rest because a velocity boost is a symmetry of the system. Yet, these are distinct physical states, for someone on the shore will notice whether the ship is at rest or moving with uniform velocity. Here we see that the velocity boost is a symmetry of the subsystem, but when the boost is applied to the subsystem alone it is no longer a symmetry of the whole world and we will be able to know that these are physically different states when we interact with any part of the world that is not within the subsystem. That is, a global symmetry applied to a subsystem generates an empirically distinct state of the world, whereas a global symmetry applied to the whole world would not generate a empirically distinct state of the world. 

On the contrary, local symmetries (recall: those which need not act identically at each point in spacetime) are not considered to be directly empirically significant because they are believed to encode mere descriptive redundancy in our physics.
Thus, the thinking goes, it is impossible to have an analogue of the Galileo's ship scenario for gauge symmetries (for literature defending and opposing this view, see \cite{BB, Friederich, SMR}).
This does not mean, however, that local symmetries are \emph{entirely} devoid of empirial content. For example, the fact that a gauge theory has these symmetries implies conservation laws important conservation laws from Noether's theorems. This suggests we can classify symmetries as having either direct or indirect empirical significance, which are defined in the following way:
\begin{description}
\item[Direct Empirical Significance (DES):] A symmetry is said to have direct empirical significance if it is characterized by the following conditions \cite{BB}: (i) the transformation of the subsystem with respect to a reference system leads to physically distinct scenarios; (ii) the transformation of the subsystem is a symmetry of the subsystem and is empirically indistinguishable from within the subsystem.

\item[Indirect Empirical Significance (IES):] A symmetry is said to have indirect empirical significance if there are empirical consequences associated with the symmetry. For example, the fact that these gauge symmetries exist places restrictions on the equations of motion, which implies that charges are conserved through Noether's theorems \cite{Gomes}. Thus, these symmetries are still empirically meaningful in the sense that there are empirical consequences to the fact that physical laws possess these symmetries, even if they don't possess DES in the sense discussed above of being able to generate physically distinct states when solely applied to subsystems. 
\end{description}

Much recent work---promulgated by Greaves and Wallace \cite{GW}---has clarified the extent to which symmetries can be empirically significant and shown that there are perhaps more similarities between these types of symmetries than the traditional view acknowledges. In particular, the present article intersects with two recent streams of thought regarding the empirical significance of symmetries: (a) the distinction between Type I and Type II DES articulated in \cite{GW}, and (b) the distinction between `artificial' and `substantial' gauge symmetry  as developed by François \cite{Francois}. In the following subsections we explore how our model of the edge modes in the Newton-Cartan QHE intersects with these classifications.

\subsection{Type I and Type II empirical significance}

Greaves and Wallace argue in \cite{GW} that gauge symmetries can indeed---contrary to the orthodoxy expressed in e.g.~\cite{BB}---possess DES in addition to IES. In order to do this, they illustrate how applying gauge symmetries to subsystems can create analogues to Galileo's ship type scenarios. Furthermore, they propose the following two schemes for illustrating the DES of a symmetry: Type I scenarios and Type II scenarios.

Type I scenarios correspond to exactly what we have discussed regarding the standard Galileo's ship type scenario. In a slightly more formal way of describing the situation, consider an ordered pair $ \braket{s,e}$, where $s$ and $e$ correspond to dynamically allowed states for a subsystem and environment that are compatible with each other. Consider also that the universe possesses a set of dynamical symmetries $\sigma$. A symmetry has Type I empirical significance when $ \braket{\sigma(s),e}$ is defined and instantiates a physically distinct state of the universe from $ \braket{s,e}$.\footnote{Here, the two states are related by the transformation $\sigma|_S * \mathbb{I}$---i.e.,~the transformation which acts as $\sigma$ on the subsystem, and as the identity on the environment.} That is, there is a relational difference between the subsystem and environment. Both \cite{Teh} and \cite{Gomes} argue that we can indeed construct analogues to the Galileo's ship scenario for Yang-Mills type gauge theories. 

Type II scenarios, on the hand, invite us to likewise consider a subsystem and environment pair,  $\braket{s,e}$, along with a set of boundary conditions for the subsystem states $C_{e}$ that is compatible with the environment states. Then imagine a symmetry $\sigma$ that is applied to one of these subsystem states such that $\sigma (C_{e}) = C_{e'}$, where $e$ and $e'$ are physically distinct environmental states representing different charged states. Here, we have $ \braket{\sigma(s),e} \rightarrow  \braket{\sigma(s),e'}$, where we can see that the application of the symmetry has created a non-trivial change in the environment state itself. Furthermore, there is understood to be a `principled connection' between the gauge symmetry and the change in environment state $e\rightarrow e'$. While the Greaves and Wallace notion of a `principled connection' is somewhat vague, as we shall see, other authors have proposed cashing out this idea in terms of the boundary charge that corresponds to this change in environmental state generating the relevant symmetry that can fulfill this role \cite{TR}. Type I scenarios change the subsystem state while keeping the environmental state the same, whereas Type II scenarios change both the subsystem and environment states.

Not surprisingly, the quantum Hall system we have considered in this paper will be interesting for the Type II scenario because it is concerned with this issue of gauge symmetries and the existence of edge modes on the boundary.  Murgueitio Ramirez and Teh consider in \cite{TR} a related scenario in which they assess the Type II empirical significance of a $U(1)$ gauge theory with boundary. In order to assess this situation, they introduce a boundary field to restore $U(1)$ gauge invariance in analogy with the Donnelly-Freidel procedure used above and indeed find a new boundary symmetry in addition to the $U(1)$ gauge symmetry in complete analogy to the new boundary symmetry that emerged in our QHE model. They contend---contrary to Greaves and Wallace's argument---that the $U(1)$ gauge symmetry of the subsystem could not be construed as having Type II empirical significance because (as indicated above) the most natural way of cashing out such a `principled connection' is the traditional relationship between symmetries and conserved charges, where the charges generate the symmetries. The subsystem symmetry $\sigma$ on this $U(1)$ theory is not generated by the boundary charges. Rather, the new boundary symmetry is generated by the boundary charges. They conclude that the Type II scenario for DES can't be correct in the case of the $U(1)$ gauge symmetry $\sigma$; however, the boundary symmetry does realize this notion of Type II empirical significance because it can affect a transition from $e \rightarrow e'$.

Coming to our model of the Newton-Cartan QHE, there are three prominent symmetries of which we have made use: the Milne symmetry, the $U(1)$ gauge symmetry, and the boundary symmetry. Recall that we dressed up the na\"{i}ve mass gauge field of the Newton-Cartan structure to be Milne invariant in order both to provide invariance for the topological theory under the relevant non-relativistic symmetries that are exhibited by the microscopic theory and are necessary to more generally model the QHE, and to avoid the conceptual difficulties of imposing non-relativistic diffeomorphism invariance on the background gauge fields. Additionally, we introduced a boundary field $u$ to make the pre-symplectic potential gauge invariant under the $U(1)$ symmetry and this allowed us to see clearly the existence of chiral boson propagating on the edge of the quantum Hall system. Finally, this construction allowed us to see a new boundary symmetry that is generated by the observables. 

It is evident that the traditional view that constraints on the allowed gauge transformations lead to the physical degrees of freedom manifested as the edge modes on the boundary is conceptually very different from the approach deployed in this article (for a clear comparison of such methodologies, see \cite[p.~44]{Geiller}).
Additionally, we find that this scheme for understanding the empirical significance of symmetries maps well onto our model and the symmetries that have been relevant in its construction:
\begin{enumerate}
\item We find that the Milne symmetry can clearly exhibit Type I DES if it is construed as a spatiotemporal symmetry, and is applied to the subsystem.
This is expected because this encodes well-known symmetries such as Galilean boosts and demanding that our system be invariant under these is a simple physical requirement for non-relativistic systems. Similarly, when applied to the subsystem, this will still be a symmetry of the subsystem and satisfy these physical requirements while also generating a distinct empirical state in relation to the rest of the universe. On the other hand, if the Milne symmetry is construed as an internal, gauge symmetry (involving rescalings of the Newton-Cartan fields, as presented in \eqref{Milne}---cf.~standard $U(1)$ gauge transformations in electromagnetism), then its empirical significance is much more delicate (and, arguably, non-existent!), as is revealed in the following two points.\footnote{Clearly, not all symmetries afford an interpretation both in terms of an external, `spatiotemporal' symmetry and in terms of an internal, `gauge' symmetry. Exploring when both such interpretations are available, and the foundational significance of this possibility, we will leave as a task for future exploration.} 
\item We can see that the new boundary symmetry exhibited by this construction exhibits Type II DES in the sense that it can lead to distinct empirical states through shifting the charges in relation to the environment, yet be empirically indistinguishable from within the subsystem. Furthermore, the boundary symmetry exhibits a `principled connection' with the change in environmental state through the fact that it is generated by the boundary observables.
\item The $U(1)$ gauge symmetry---once featuring so prominently in this story of QHE edge modes (as it was believed that constraints on these transformations represented the physical degrees of freedom seen in edge modes)---is still present, but is not directly responsible for the empirical consequences of the symmetries of the system. While it still plays a role in the formalism of the model through the introduction of the field that restores the $U(1)$ invariance of the theory and corresponds to the chiral boson on the edge of the system, it does not have the direct empirical consequences that much of the previous literature had ascribed to it.
\end{enumerate}
\subsection{Artificial and substantial gauge symmetries from the dressing field method}

Another line of thought regarding the empirical content of symmetries that has emerged recently is the distinction between `artificial' and `substantial' gauge symmetries and the idea that this can be seen through applying the `dressing field method' \cite{Francois}. Here an `artificial' symmetry is one which simply represents surplus structure or descriptive reduncancy, without having direct physical signatures. A `substantial' symmetry, on the other hand, is one that does encode real physical degrees of freedom that can't be dispensed without trading off the non-locality of the theory. 
We have relied on the dressing field method extensively in this paper in constructing the Newton-Cartan QHE model and in extracting the boundary observables. We both dressed the mass gauge field to be invariant under Milne transformations and dressed the pre-symplectic potential to be invariant under $U(1)$ gauge transformations on the boundary. What do these imply about the empirical content of these symmetries and can we understand them as being `artificial' or `substantial'?

Following \cite{Francois}, consider the familiar fields $A$, $F$, and $\varphi$ where  $A$ is a gauge potential, $F$ is the field strength of the gauge potential, and $\varphi$ is a matter field. Furthermore, consider how the gauge transformation $\gamma$ (with a Lie group $H$) acts on the fields and covariant derivative $D$. In typical theories such as the Klein-Gordon or Yang-Mills Lagrangians, one finds that their Lagrangians are gauge invariant under these gauge transformations:
\begin{equation}
\begin{aligned}
A^{\gamma} &=\gamma^{-1} A \gamma+\gamma^{-1} d \gamma, & F^{\gamma}=\gamma^{-1} F \gamma, \\
\varphi^{\gamma} &=\rho\left(\gamma^{-1}\right) \varphi, \quad  & D \varphi^{\gamma}=\rho\left(\gamma^{-1}\right) D \varphi.
\end{aligned}
\end{equation}

The dressing field method comes into play by supposing that $H$ has some subgroup where there is a field $u$ that transforms as $u^{\gamma} = \gamma^{-1} u$. We can then dress the original fields with this new field $u$ and form the dressing fields:

\begin{equation}
\begin{aligned}
A^{u} &:=u^{-1} A u+u^{-1} d u, & F^{u}:=u^{-1} F u, \\
\varphi^{u} &:=\rho\left(u^{-1}\right) \varphi, \quad  & D^{u} \varphi^{u}:=\rho\left(u^{-1}\right) D \varphi.
\end{aligned}
\end{equation}

We see that, contrary to the original gauge-variant variables which transform under the gauge transformation $\gamma$, these dressed fields are now gauge-invariant variables. That is, we have effectively erased the gauge symmetry in the Lagrangian because the new fields themselves are already manifestly gauge invariant. There is no longer a symmetry, there are just the dressed variables. This is exactly what we did in dressing the Newton-Cartan mass gauge field. That is, $\overset{I}{a}$ is manifestly invariant under all Milne boosts and a gauge-invariant field. We did something slightly different in introducing our $u$ on the boundary. We introduced the dressing field $u$, which transforms in such a way that $U(1)$ gauge transformations are a symmetry of the pre-symplectic potential. That is, these are still gauge-variant variables. However, we could just as easily and equivalently introduced $\overset{I}{a^u} = \overset{I}{a} + du$, which transforms as $\overset{I}{a^u} \rightarrow \overset{I}{a} +d\alpha + du -d\alpha = \overset{I}{a^u} $ and is a gauge-invariant field. Here again, it seems like we can equivalently work in variables where the symmetry is evident or variables which do not transform under the symmetry. 

François argues that the ability to find local dressing fields in this way is a powerful indicator that the gauge symmetry is artificial. That is, the gauge-invariant variables represent the true physical fields and the symmetries correspond to these descriptive redundancies, which can be kept or dismissed without any consequences. On the contrary, if the introduction of a dressing field requires that the dressing field be non-local, as is the case in the example of Dirac's gauge-invariant reformulation of electromagnetism (see \cite{Francois}), this means that there is a price to pay for erasing the gauge symmetry in this way. Such a symmetry is considered to be substantial because it is said to probe genuine physical content. Again this does not preclude there from being some sense of IES in the sense of Noether charges, but this view holds that the symmetries themselves don't have DES, but nevertheless reveal descriptive redundancy that can have indirect significance.  

How does this artificial/substantial distinction map onto the symmetries of our Newton-Cartan QHE model? Here, our answers are as follows:
\begin{enumerate}
\item This distinction maps quite well onto the discussion of DES and IES for the case of the $U(1)$ symmetry. In the previous discussion, we found that the DES of the model was contained in the boundary symmetry because it is generated by the boundary charges and its application can lead to physically distinct states in the environment. On the contrary, the $U(1)$ symmetry was not found to have DES. Similarly, with this framework we can freely dress away the $U(1)$ of the pre-symplectic potential and find that this is an artificial symmetry.
\item The case of Milne invariance is a little more involved. If Milne invariance is construed as a local gauge transformation, this amounts to a transformation of the Newton-Cartan structures and fields such as $h^{\mu\nu}$, $n_\mu$, and $\overset{I}{a}_\mu$, and would seem to indicate that the symmetry is artificial in the sense that we can erase it by dressing it away as we did with the $U(1)$ symmetry. However, Milne symmetry can also be construed as an `external' symmetry and applied to a subsystem, as in the case of Galileo's ship. In this case we saw earlier that there is a clear case where it exhibits Type I DES. Yet, we can still dress away the symmetry. This seems to indicate that finding a dressing field does not universally indicate that there can be no DES to the symmetry in question---contrary to the claims made in \cite{Francois}.
\end{enumerate}

\section{Conclusion}

This work has constituted a further exploration into the application of Newton-Cartan theory to the study of non-relativistic systems, in particular the quantum Hall effect.
Specifically, we have applied the Donnelly-Freidel program to the bulk model of the Newton-Cartan QHE in order to construct edge modes for the model. We have shown that one can describe the edge modes of the Newton-Cartan QHE by working with a model that both (i) respects the Milne symmetry of the Newton-Cartan structure that non-relativistic systems must obey, as well as (ii) satisfies the requirements for $U(1)$ gauge invariance even in the presence of a boundary through the introduction of an additional boundary field that does not require any specific gauge fixing procedure. This construction offers key insights into the origins of these edge modes, as well as into the nature of the empirical significance of the symmetries relevant to this model, in particular the Milne symmetry, the $U(1)$ gauge symmetry, and the new boundary symmetry that came into focus after enhancing the model through the addition of a new boundary field. It is our hope that, in turn, such methodologies will help to deliver further insights into \emph{\`{a} la mode} contemporary topics, such as non-relativistic holography (on which see e.g.~\cite{C1, C2}), or edge modes in gravitational theories (on which see e.g.~\cite{L1, L2}).

\appendix

\acknowledgments

We are very grateful to Sebastián Murgueitio Ramirez and to the anonymous referee for very helpful comments on a previous draft of this article. We also thank Marc Geiller, Henrique Gomes, Andrey Gromov, Carlos Hoyos, Philippe Mathieu, and Laura Murray for helpful discussions. This work was supported by the John Templeton Foundation, grant number 61521, and the National Science Foundation, grant number 1947155.




\begin{thebibliography}{99}




\bibitem{Avron} J. E. Avron, R. Seiler, and P. G. Zograf, ``Viscosity of Quantum Hall Fluids'', Physical Review Letters 75, pp.~697-700, 1995.

\bibitem{Ban1} Rabin Banerjee, Pradip Mukherjee, ``Milne Boost from Galilean Gauge Theory'', Physics Letters B 778, pp.~303-308, 2018.

\bibitem{BB} Katherine Brading and Harvey R.~Brown, ``Are Gauge Symmetry Transformations Observable?'', British Journal for the Philosophy of Science 55, pp.~645–665, 2004.

\bibitem{C1} M.~H.~Christensen, J.~Hartong, N.~A.~Obers, and B.~Rollier, ``Torsional Newton–Cartan Geometry
and Lifshitz Holography'',  Physical Review D 89 061901, 2014.

\bibitem{C2} M.~H.~Christensen, J.~Hartong, N.~A.~Obers, and B.~Rollier, ``Boundary Stress–energy Tensor and
Newton–Cartan Geometry in Lifshitz Holography'' Journal of High Energy Physics, JHEP(2014)57, 2014.

\bibitem{DL}
William Donnelly and Laurent Freidel, ``Local Subsystems in Gauge Theory and Gravity'', Journal of High Energy Physics, JHEP(2016)102, 2016.

\bibitem{Francois} Jordan François, ``Artificial versus Substantial Gauge Symmetries: A Criterion and an Application to the Electroweak Model'', Philosophy of Science 86, pp.~472-496, 2019.

\bibitem{L1} Laurent Freidel, Marc Geiller and Daniele Pranzetti, ``Edge Modes of Gravity. Part I. Corner Potentials and Charges'', Journal of High Energy Physics, 2020.

\bibitem{L2} Laurent Freidel, Marc Geiller and Daniele Pranzetti, ``Edge Modes of Gravity. Part II. Corner Metric and Lorentz Charges'', Journal of High Energy Physics, 2020.

\bibitem{Geiller}
Marc Geiller, ``Edge Modes and Corner Ambiguities in 3D Chern-Simons Theory and Gravity'', Nuclear Physics B 924, pp.~312-365, 2017.

\bibitem{GeraciePrabhu}
Michael Geracie, Kartik Prabhu, Matthew M. Roberts, ``Covariant Effective Action for a Galilean Invariant Quantum Hall System'', Journal of High Energy Physics, JHEP(2016)92, 2016.

\bibitem{Gomes}
Henrique Gomes, ``Holism as the Significance of Gauge Symmetries'', arXiv:1910.05330v2, 2020.

\bibitem{GW} Hilary Greaves and David Wallace, ``Empirical Consequences of Symmetries'', British Journal for the Philosophy of Science 65, pp.~59-89, 2014.

\bibitem{Gromov} Andrey Gromov, Kristan Jensen, and Alexander G. Abanov, ``Boundary Effective Action for Quantum Hall States'', Physical Review Letters 116, 126802, 2016.

\bibitem{DamSon2}
Carlos Hoyos, Dam Thanh Son, ``Hall Viscosity and Electromagnetic Response'', Physical Review Letters 108, 066805, 2012.

\bibitem{MorozReply} J.~Höller and N.~Read, ``Comment on ``Galilean Invariance at Quantum Hall Edge'', Physical Review B 93, 197401, 2016.

\bibitem{Jensen1}
Kristan Jensen, ``On the Coupling of Galilean-Invariant Field Theories to Curved Spacetime'', SciPost Physics 5, 011, 2018.

\bibitem{Jensen2}
Kristan Jensen, ``Aspects of Hot Galilean Field Theory'', Journal of High Energy Physics, JHEP(2015)123, 2015.

\bibitem{Malament} David B.~Malament, \emph{Topics in the Foundations of General Relativity and Newtonian Gravitation Theory}, Chicago, IL:~Chicago University Press, 2012.

\bibitem{LMP} Philippe Mathieu, Laura Murray, Alexander Schenkel and Nicholas J.~Teh, ``Homological Perspective on Edge Modes in Linear Yang-Mills and Chern-Simons Theory'', Letters in Mathematical Physics 110, pp.~1559–1584, 2020.

\bibitem{Hoyos}
Sergej Moroz, Carlos Hoyos and Leo Radzihovsky, ``Galilean Invariance at Quantum Hall Edge'',
Physical Review B 91, 195409, 2015.

\bibitem{MorozCorrection} Sergej Moroz, Carlos Hoyos, and Leo Radzihovsky, ``Erratum: Galilean Invariance at Quantum Hall Edge'', Physical Review B 96, 039902, 2017.

\bibitem{TR} Sebastián Murgueitio Ramirez and Nicholas J.~Teh, ``Abandoning Galileo's Ship: The Quest for Non-relational
Empirical Significance'', British Journal for the Philosophy of Science, forthcoming.

\bibitem{TehRead}
James Read and Nicholas J.~Teh, ``The Teleparallel Equivalent of Newton-Cartan Gravity'', Classical and Quantum Gravity 35, 18LT01, 2018.

\bibitem{ReadRezayi} N.~Read and E.~H.~Rezayi, ``Hall Viscosity, Orbital Spin, and Geometry: Paired Superfluids and Quantum Hall Systems'', Physical Review B 84, 085316, 2011.

\bibitem{DamSon1}
Dam Thanh Son, ``Newton-Cartan Geometry and the Quantum Hall Effect'',
arXiv:1306.0638, 2013.

\bibitem{DamSon3}
D.~T.~Son, M.~Wingate, ``General Coordinate Invariance and Conformal Invariance in Nonrelativistic Physics: Unitary Fermi Gas'', Annals of Physics 321(1), pp.~197-224, 2006.

\bibitem{Teh} Nicholas J.~Teh, ``Galileo's Gauge: Understanding the Empirical Significance of Gauge Symmetry'', Philosophy of Science 83, pp.~93-118, 2016.

\bibitem{TehNewton} Nicholas J.~Teh, ``Recovering Recovery: On the Relationship between Gauge Symmetry and Trautman Recovery'', Philosophy of Science 85(2), pp.~201-224, 2018.

\bibitem{Tong} David Tong, ``The Quantum Hall Effect'', TIFR Infosys Lectures, 2016.

\bibitem{Wen}
Xiao-Gang Wen, ``Theory of the Edge States in Fractional Quantum Hall Effects'',
International Journal of Modern Physics B 6(10), pp.~1711-1762, 1992.

\bibitem{Zirn}
Martin R. Zirnbauer, ``The Integer Quantum Hall Plateau Transition is a Current Algebra After All'', Nuclear Physics B 941, pp.~458-506, 2019.

\bibitem{Friederich}
Simon Friederich, ``Symmetry, empirical equivalence, and identity'',
The British Journal for the Philosophy of Science
Vol. 66, No. 3, pp. 537-559, 2015.

\bibitem{SMR}
Sebastián Murgueitio Ramírez, ``A puzzle concerning local symmetries and their empirical significance'', The British Journal for the Philosophy of Science (forthcoming).












\end{thebibliography}
\end{document}